\documentclass[sigplan, 10pt, screen]{acmart}

\usepackage[utf8]{inputenc}
\usepackage[T1]{fontenc}
\usepackage{hyperref}
\usepackage{url}
\usepackage{booktabs}
\usepackage{nicefrac}
\usepackage{microtype}
\usepackage{xcolor}
\usepackage{listings}
\usepackage{multirow}
\usepackage{threeparttable}

\usepackage{arydshln}
\usepackage[textsize=tiny,textwidth=0.6in]{todonotes}
\newcommand{\allnotes}[1]{}
\renewcommand{\allnotes}[1]{#1} 

\setcopyright{acmcopyright}
\copyrightyear{2026}
\acmYear{2026}
\acmDOI{XXXXXXX.XXXXXXX}

\acmConference[EuroMLSys]{The 6th Workshop on Machine Learning and Systems}{April 2026}{Edinburgh, United Kingdom}
\acmBooktitle{Proceedings of the 6th Workshop on Machine Learning and Systems}
\acmPrice{15.00}
\acmISBN{978-1-4503-XXXX-X/18/06}

\title{Pooling Engram Conditional Memory in Large Language Models using CXL}

\author{Ruiyang Ma$^{1,\ast}$\char44\ Teng Ma$^{2}$\char44\ Zhiyuan Su$^{3}$\char44\ Hantian Zha$^{4,\ast}$\char44\ Xinpeng Zhao$^{2}$\char44\ Xuchun Shang$^{2}$\char44\ Xingrui Yi$^{2}$\char44\ Zheng Liu$^{2}$\char44\ Zhu Cao$^{3}$\char44\ An Wu$^{3}$\char44\ Zhichong Dou$^{3}$\char44\ Ziqian Liu$^{5}$\char44\ Daikang Kuang$^{1}$\char44\ Guojie Luo$^{1}$}

\affiliation{%
  \institution{
    $^1$Peking University \quad 
    $^2$Alibaba Cloud \quad 
    $^3$Shandong Yingxin Computer Technology Co., Ltd \quad \\
    $^4$Renmin University of China \quad 
    $^5$The University of Hong Kong
  }
  \country{}
}

\thanks{$^\ast$This work was done while the author was an intern at Alibaba Cloud.}

\begin{abstract}

Engram conditional memory has emerged as a promising component for LLMs by decoupling static knowledge lookup from dynamic computation. Since Engram exhibits sparse access patterns and supports prefetching, its massive embedding tables are well-suited for offloading to lower-tier memory.
In this paper, we propose using Compute Express Link (CXL) memory pool for Engram storage. Compared to RDMA, CXL provides fine-grained and low-latency access required by minimal and discrete retrieval patterns of Engram. We integrate the CXL-based Engram pool into SGLang, achieving near-DRAM end-to-end performance. This provides a scalable and cost-efficient storage solution for future Engram-integrated LLMs without compromising inference performance.

\end{abstract}


\begin{document}

\maketitle

\section{Introduction}
Large Language Models (LLM) primarily rely on Mixture-of-Experts (MoE) for conditional computation~\cite{liu2024deepseek, liu2025deepseek}. However, this architecture lacks a native mechanism for efficient vocabulary knowledge lookup, forcing the model to simulate retrieval through computation inefficiently.
To address this, Engram introduces conditional memory as a complementary axis of sparsity~\cite{cheng2026conditional}. By modernizing N-Gram embeddings into a scalable $O(1)$ lookup mechanism, Engram decouples knowledge storage from active computation. This architecture exhibits sustained scaling laws, where performance improves consistently as the N-gram table capacity increases exponentially, making it potentially a foundational component in next-generation large LLMs.

The emergence of Engram introduces substantial memory overhead, potentially scaling to hundreds of GBs for future LLMs. Despite this footprint, the Engram module exhibits a sparse and minimal memory access pattern. Furthermore, because the retrieval embeddings can be prefetched to overlap with the computation of non-Engram layers, the module is relatively latency-tolerant. These characteristics make Engram parameters highly compatible with offloading into lower-tier memory. Consequently, a shared Engram memory pool offers a viable solution to mitigate storage overhead without compromising inference throughput.

In this paper, we demonstrate the feasibility of using CXL pooling for Engram memory management, encompassing both cost-efficient storage and high-performance retrieval.
CXL enables the construction of a disaggregated, host-independent memory pool through a CXL switch~\cite{xconn_apollo_product}, 
allowing multiple compute nodes to share a centralized memory space~\cite{das2024introduction}. Unlike RDMA-based pooling, CXL provides hardware-level support for native load/store primitives. By bypassing the overhead of traditional networking stacks, CXL ensures the low-latency, fine-grained data access~\cite{yang2025beluga} required by Engram’s sparse and discrete retrieval patterns.

We offload Engram parameters to a CXL-based memory pool and develop specialized CXL access routines optimized for high-concurrency Engram embedding transfers. 
We evaluate the performance of CXL pooling by benchmarking raw Engram embedding retrieval and end-to-end throughput within the state-of-the-art inference framework SGLang~\cite{sglang}. Our results demonstrate that CXL-based memory pooling effectively supports Engram’s conditional memory, underscoring its potential as a cost-efficient and scalable infrastructure for next-generation, memory-augmented LLMs.

In summary, the contributions of our paper are as follows:
\begin{itemize}
  \item We propose the first system to offload Engram parameters on CXL-based memory pool.
  \item We provide analysis of RDMA and CXL for Engram embedding retrieval, demonstrating the superiority of CXL pooling for Engram's memory access pattern.
  \item We implement CXL-based Engram pool on SGLang and achieve near-DRAM performance, proving the feasibility of CXL pooling for Engram.
\end{itemize}
\section{Background}
\subsection{Engram Conditional Memory}
The Engram architecture is proposed by DeepSeek to address the inefficiency in traditional MoE models~\cite{cheng2026conditional}.
By forcing Transformer networks to encode static knowledge within parameters, existing MoE models waste significant compute on simple vocabulary matching. 
Engram introduces the concept of \textit{Conditional Memory}, a complement to the \textit{Conditional Computation} in Mixture-of-Experts (MoE), to decouple static N-Gram knowledge lookup from dynamic reasoning.

\begin{figure}
    \centering
    \includegraphics[width=0.8\linewidth]{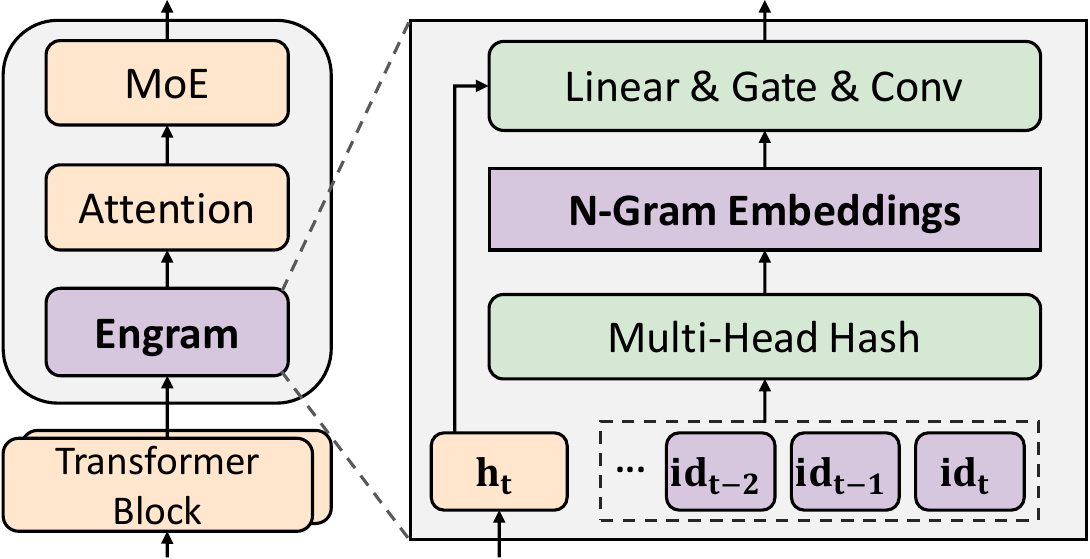}
    \caption{Brief architecture of Engram.}
    \vspace{-10pt}
    \label{fig:engram}
\end{figure}

Figure~\ref{fig:engram} illustrates the architecture of Engram-integrated language models. The Engram module is inserted into specific Transformer layers (e.g., layers 2 and 15 of 36), positioned immediately before the attention blocks.
For each token $t$ in the sequence, the module first extracts multi-granular N-Grams (e.g., N=2, N=3) and maps the token IDs to specific indices using a multi-head hashing function. These N-Gram embeddings are then asynchronously fetched and projected back into the model's dimension. Finally, a gating mechanism fuses these retrieved embeddings with the Transformer's current hidden states $h_t$, adaptively modulating the injection of retrieved knowledge based on the context. 

This design produces a distinctive \textit{storage-intensive yet compute-sparse} workload profile. Since the vast majority of parameters (specifically the N-Gram embeddings) are accessed via simple lookups based on hashed token IDs, the architecture facilitates efficient prefetching and offloading to lower-tier storage. Therefore, Engram is well-suited for memory pooling, which provides a solution to overcome the GPU memory wall and reduce hardware costs.

\subsection{Memory Pooling in LLM Serving}
\begin{figure}
    \centering
    \includegraphics[width=1.0\linewidth]{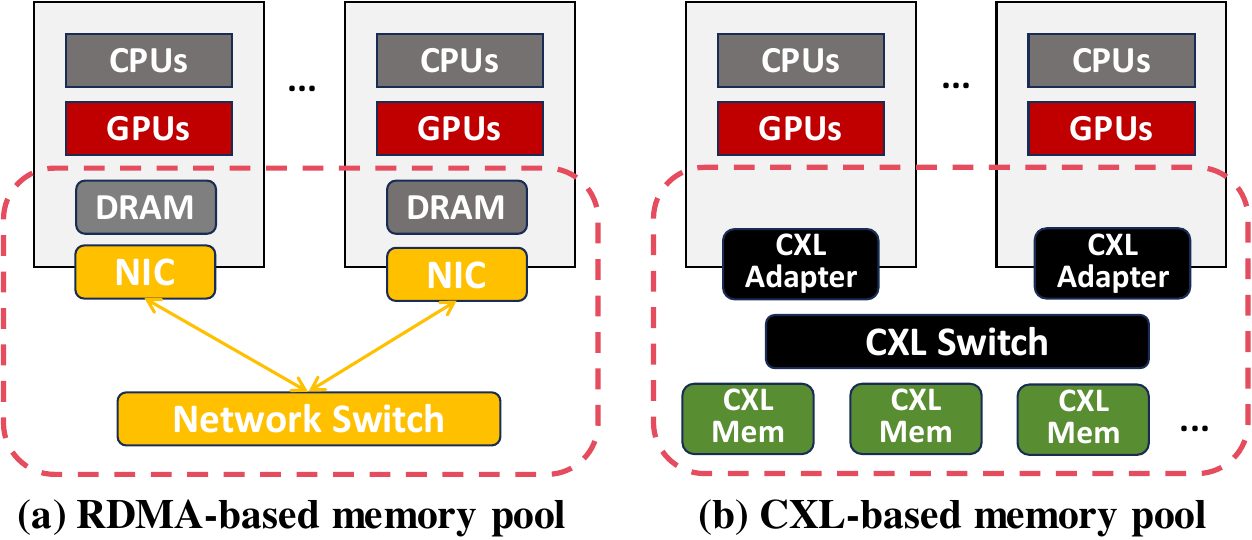}
    \caption{Overview of RDMA/CXL memory pools.}
    \vspace{-10pt}
    \label{fig:pool}
\end{figure}
The rapid scaling of LLM parameters and the escalating demand for long-context inference have made memory a critical bottleneck in GPU-accelerated serving systems, especially for the storage of KV Cache. To address this, the industry is shifting toward remote resource pooling infrastructures to achieve higher throughput and improved cost-efficiency~\cite{kvcache}.

Recent memory pooling solutions in LLM serving, such as MoonCake~\cite{mooncake} and LMCache~\cite{lmcache}, have pioneered the transition by leveraging RDMA networks to construct large-scale disaggregated memory pools, as shown in Figure~\ref{fig:pool}(a). 
These systems use a CPU-driven RDMA access model to fetch remote data. The host CPU orchestrates the data transfers by moving the data from the GPU into an intermediate bounce buffer located in the host DRAM before finally issuing the RDMA request to send it to the remote pool.

While this RDMA-centric approach successfully extends memory capacity and facilitates prefix sharing across multi-server clusters, it introduces significant trade-offs. The RDMA protocol stack and message-based networking semantics lead to significant performance degradation during small packet transfers~\cite{yang2025beluga}; for 64 byte messages, RDMA throughput often plummets to less than 25\% of peak bandwidth~\cite{kalia2019datacenter}. 
To overcome these limitations, Compute Express Link (CXL) has emerged as a compelling alternative for memory pooling, which will be detailed in the next section.

\subsection{Compute Express Link (CXL)}

CXL is an emerging interconnect protocol that facilitates high-bandwidth, low-latency memory sharing between hosts and accelerators~\cite{das2024introduction}. Early CXL 1.0 implementations focused on single-node memory expansion via the CXL.mem interface. The recent commercial CXL 2.0/3.0 switches~\cite{xconn_apollo_product} 
has shifted the paradigm toward memory sharing. This evolution allows multiple host nodes to dynamically share a centralized memory pool, providing a scalable hardware foundation for managing memory-intensive workloads.

As illustrated in Figure~\ref{fig:pool}(b), compared to traditional RDMA, CXL simplifies the data path and provides a hardware-managed load/store semantic memory interface. By enabling access at cache-line granularity, CXL is particularly advantageous for the fine-grained, sparse memory lookup, achieving a low-latency remote memory access that approaches the performance of local DRAM. 

Due to these characteristics, CXL is emerging as a compelling alternative to RDMA for memory pooling across diverse data center workloads, including databases~\cite{cxl_db_1, cxl_db_2, cxl_db_3} and cloud platforms~\cite{cxl_cloud_1, cxl_cloud_2, cxl_cloud_3}. Recent research also highlights its promise in LLM, including KV cache management~\cite{yang2025beluga, yoon2025tract} and retrieval-augmented generation (RAG) systems~\cite{quinn2025accelerating, kim2026bauhaus}. 
In these established scenarios, such as the retrieval of prefix KV caches or RAG block embeddings, data movement typically occurs at the beginning of a request and involves bulk transfers ranging from the MB to GB level. In contrast, Engram introduces a more fragmented access pattern, requiring frequent, KB-level transfers across each Engram-layer of the model. In the following section, we analyze this sparse access profile and demonstrate the superiority of CXL.

\section{Analysis and Profiling}

\subsection{Memory Access Patterns of Engram}

Engram conditional memory introduces a unique access pattern that distinguishes it from traditional LLM components like KV caches or expert weights. It is characterized by three primary features:
\begin{itemize}
    
    \item \textbf{Read-only and Minimal Access:} 
    Unlike the KV cache, which must be updated and appended at every decoding step, Engram memory is static and immutable during inference. While model weights are also read-only, Engram memory is distinct in that only a minimal fraction of its parameters is retrieved per forward pass, significantly smaller than that of dense or MoE parameters loading.
    
    \item \textbf{Sparse Retrieval:} In current configurations, the model fetches only ~5 KB of data per token per layer. This payload consists of 16 discrete, small-scale embeddings (based on $N\text{-gram}=2$ and $8$ hash heads) among massive Engram table. In batched inference, the total volume and frequency of memory accesses scale linearly with the batch size. 

    \item \textbf{Latency Tolerance:} Engram modules are strategically placed in only a few transformer layers (e.g., layers 2 and 15). Since the hash function relies on N-gram token IDs rather than intermediate hidden states, retrieval can be initiated at the start of the decoding step. This allows the memory read latency to be overlapped with the execution of preceding transformer blocks.
\end{itemize}

These characteristics allow Engram to be offloaded to cost-effective, shared memory pools across nodes. However, the system must be engineered to handle sparse access patterns without exceeding the pre-layer retrieval deadline. We will analyze those requirements in the following sections. 

\subsection{Engram Requirements for Memory Pool}
Engram retrievals are triggered at specific execution stages, the memory traffic exhibits highly bursty patterns.
To ensure the offloading of Engram to memory pool does not bottleneck the existing LLM inference performance, we consider that from both bandwidth and latency constraints. 

\textbf{Bandwidth Requirement: } The average required read bandwidth $B_{pool}$ is determined by token processing rate in system. Let $T$ represent the target system throughput (tokens per second), $S_{layer}$ denote the data size of Engram embeddings retrieved per token for a single layer, and $N_{eng}$ be the total number of transformer layers that incorporate Engram modules. To maintain the baseline generation speed, the average bandwidth of the memory pool must satisfy:
\begin{equation}
    B_{pool} > T \times S_{layer} \times N_{eng} \nonumber
\end{equation}

\textbf{Latency Requirement: } The retrieval process must be completed within a \textit{prefetch window} to hide memory access overhead. Let $L_{pool}(N_{token}, S_{layer})$ be the end-to-end latency to retrieve the required Engram embeddings for a single layer. $N_{token}$ represents the number of tokens concurrently triggering retrieval in the current inference step. $t_{exec}(i)$ is the computation time of the $i$-th layer, and the Engram module is integrated at layer $k$. To ensure zero-stall execution, this latency must satisfy:
\begin{equation}
    L_{pool}(N_{token}, S_{layer}) < \sum_{i=1}^{k-1} t_{exec}(i) \nonumber
\end{equation}

\textbf{Case Study: } 
We analyze the performance of Qwen3-32B~\cite{qwen3technicalreport} using SGLang~\cite{sglang} on a $4\times$ NVIDIA H200 GPU cluster. We select Qwen3-32B because its parameter scale and sparse architecture align with current Engram configurations. It provides an open-source alternative, while the weight for DeepSeek Engram remains undisclosed. 
Table~\ref{tab:case} summarizes the basic parameters in this case.

\begin{table}[h]
\centering
\caption{System configuration and experimental parameters.}
\vspace{-5pt}
\label{tab:case}
\begin{tabular}{lcc}
\hline
\textbf{Parameter} & \textbf{Symbol} & \textbf{Value} \\ \hline
System Throughput    & $T$         & $\approx$ 70,000 tokens/s \\
Decode Step Latency  & $t_{step}$  & $\approx$ 3.6 ms \\
Engram Modules       & $N_{eng}$   & 2 layers ($k=2, 15$) \\
Retrieval Payload    & $S_{layer}$ & 5 KB / token / layer \\
Batch Size           & $N_{token}$ & 256 \\ \hline
\end{tabular}
\end{table}

\textit{Bandwidth Constraint:} At a throughput of 70k tokens/s, the total required read bandwidth is $B_{pool} = T \times S_{layer} \times N_{eng} \approx \mathbf{0.7\text{ GB/s}}$. It is easily accommodated by PCIe Gen5 (64 GB/s) or 100GbE network (12.5 GB/s). Thus, bandwidth is not the primary bottleneck for Engram offloading.

\textit{Latency Constraint:} Given $t_{step}\approx3.6\text{ms}$ across 64 layers, the average computation time per layer ($t_{exec}$) is approximately 56$\mu$s. For an Engram module at Layer 2, the prefetch window in the system is restricted to about 56$\mu$s. This is a stringent requirement for pooled memory systems.

Our analysis shows that while the memory pool easily satisfies bandwidth requirements, the prefetch window imposes relatively strict latency constraints, particularly for early-layer modules. In the following section, we investigate whether memory pools can meet these requirements.

\subsection{Profiling Pooling Fabrics: RDMA and CXL}
\label{sec:rdma_cxl}
\begin{figure}[h]
    \centering
    \includegraphics[width=1.0\linewidth]{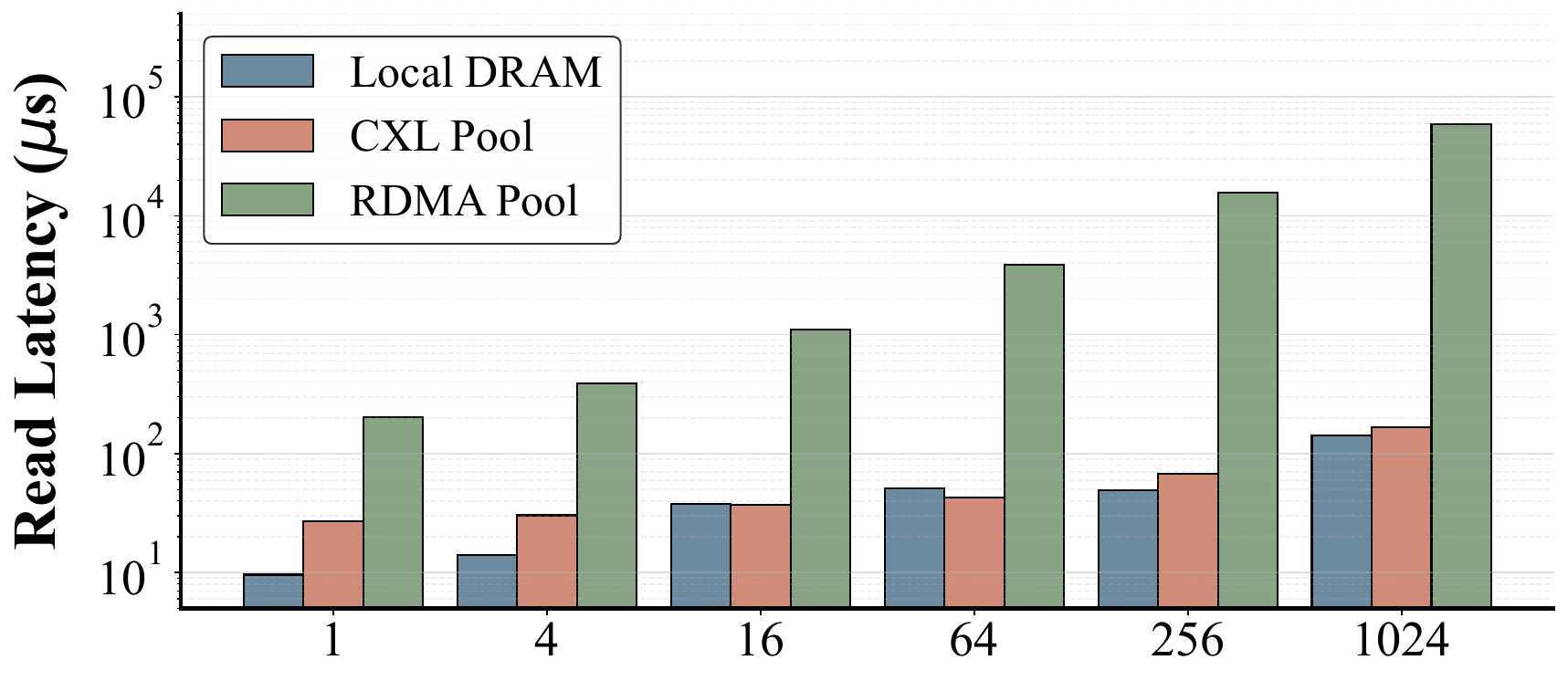}
    \vspace{-20pt}
    \caption{Latency for Engram-27B across varying batch size.}
    \vspace{-10pt}
    \label{fig:cxl_rdma}
\end{figure}

We profile two dominant memory pooling architectures for Engram embedding storage: RDMA and CXL. Our experimental setup utilizes a two-server configuration. Detailed configuration are provided in Section~\ref{sec:exp_set}.
\begin{itemize}
    \item RDMA Pooling: Implemented via Mooncake Store~\cite{mooncake}, where Engram embeddings are partitioned equally across the DRAM of both servers. Data is retrieved using standard \texttt{put} and \texttt{get} operations.
    \item CXL Pooling: Implemented using a CXL Switch connected to a CXL memory card, allowing shared access from both servers.
\end{itemize}
We evaluate the latency of reading Engram embeddings from the pool to the CPU. Following the Engram-27B configuration ($\text{vocab\_size}=2,262,400$; $\text{emb\_dim}=1,280$), each token retrieval requires fetching 8 hash-mapped 320 bytes data segments, distributed across discrete, sparse addresses. Figure~\ref{fig:cxl_rdma} compares the latency across varying batch sizes of retrievals.



The profiling results indicate that RDMA incurs a latency penalty orders of magnitude higher than that of local DRAM. In contrast, CXL-based memory pooling delivers latencies that closely approximate local DRAM, aligning well with the latency timing requirements of Engram prefetching. 

\section{Methodology}
\subsection{Overview}
\begin{figure}
    \centering
    \includegraphics[width=1.0\linewidth]{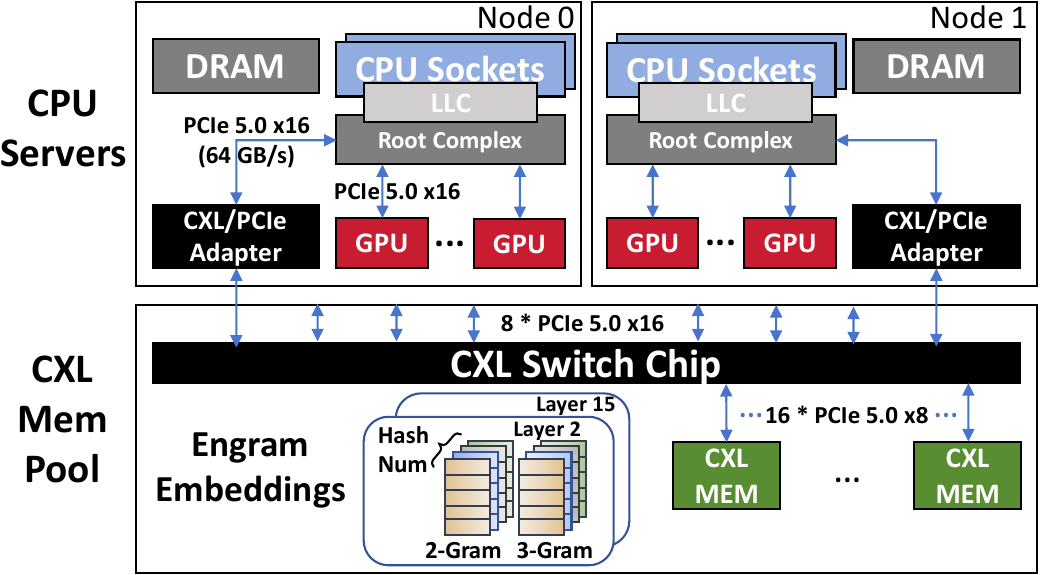}
    \vspace{-10pt}
    \caption{Overview of CXL-based Engram pooling system.}
    \vspace{-10pt}
    \label{fig:overview}
\end{figure}

Figure~\ref{fig:overview} illustrates the overview of CXL-based Engram pooling system. Each server in the cluster contains host CPUs and multiple GPUs. Each NUMA node is connected to the CXL switch via a PCIe 5.0 x16 PCIe/CXL adapter.
The CXL memory pool consists of a switch node, which is equipped with XConn XC50256 CXL switch chip~\cite{xconn_apollo_product}. The chip has 256 PCIe 5.0 lanes which are partitioned between CXL memory devices and compute servers. The CXL switch chip can provide 512 GB/s of total bandwidth between memory devices and compute servers, which connects up to 8 servers to a 4 TB memory pool. This connectivity enables the CXL pool to support concurrent multi-host access through its internal address mapping and forwarding logic.

In our prototype, since the footprint of the Engram conditional memory in current models is relatively small, we store the entire Engram table within a single, contiguous 256 GB CXL memory card. This centralized Engram table is shared across multiple servers, allowing both CPUs and GPUs to perform direct memory operations through the CXL switch. 

\subsection{CXL Access Routines}

CXL shared memory is exposed as a DAX (Direct Access) device in operating system. To initialize the memory segment, a standard \texttt{mmap} system call maps the DAX device to a virtual address pointer. Once mapped, the CXL memory space can be read from or written to as a standard data segment within the application, allowing for seamless integration with existing memory management workflows. Since the Engram embeddings are read-only, the system avoid the overhead of maintaining cache coherence across nodes.


During each Engram retrieval, CPU or GPU must fetch a large volume of small, discrete embeddings, typically thousands of segments, each hundreds of bytes in size. To fully saturate the PCIe bandwidth, it is critical to maximize the parallelism of these data transfers. To address this, we implemented highly efficient CXL-read functions in C++ for both the CPU and GPU, optimized to handle high-concurrency memory requests and minimize transmission overhead.

\textbf{CXL$\to$CPU:} The retrieval process from CXL to CPU employs a multi-threaded parallel read strategy based on OpenMP. Since the CXL device is mapped into the user-space virtual address space via DAX mode, the CPU can perform data transfers using standard `memcpy` as if it were accessing local DRAM.

\begin{lstlisting}[language=C++, caption={CXL to CPU Engram embedding retrieval.}]
#pragma omp parallel for schedule(static) num_threads(64)
for(int i = 0; i < count; ++i) {
    void *cxl_src = cxl_base + offsets[i];
    void *cpu_dst = cpu_base + bytes * i;
    std::memcpy(cpu_dst, cxl_src, bytes);
}
\end{lstlisting}

\textbf{CXL$\to$GPU:} For data movement from CXL to GPU, we implement a custom CUDA kernel to enable direct P2P (Peer-to-Peer) data transfer, completely bypassing the CPU after the initial memory mapping. The CXL memory region is registered as CUDA host memory during system initialization using \texttt{cudaHostRegister}. This allows the GPU's DMA engine to address CXL memory pointers directly within the kernel function via the PCIe bus. To avoid the prohibitive launch overhead of thousands of discrete \texttt{cudaMemcpy} calls, we fuse all retrieval operations into a single wide-grid kernel, which allows the GPU scheduler to saturate PCIe bandwidth by overlapping thousands of concurrent memory requests.

\begin{lstlisting}[language=C++, caption={CXL to GPU Engram embedding retrieval.}]
// custom `cxl2vram_copy` CUDA kernel 
void *cxl_src = cxl_base + offsets[blockIdx.x];
void *gpu_dst = gpu_base + blockIdx.x * bytes;
for(int i = blockIdx.x; i < bytes; i+=blockDim.x)
    gpu_dst[i] = cxl_src[i];
    
// launch the kernel
const int threads = 256;
const dim3 grid(count);
cxl2vram_copy<<<grid, threads, 0, cuda_stream>>>(...)

\end{lstlisting}

\subsection{Implementation in Inference Framework}
To integrate the CXL-based Engram memory pool into LLM inference frameworks such as SGLang~\cite{sglang}, three primary components require modification:

\textbf{Initialization:} SGLang instantiates a \texttt{ModelRunner} per rank. Within each node, the lowest-ranking \texttt{ModelRunner} is designated to configure the local CXL memory space. Only a single rank globally (e.g., \texttt{tp\_rank=0}, \texttt{pp\_rank=0}) loads the Engram parameters into the shared CXL memory pool.

\textbf{Prefetching:} Upon the arrival of a \texttt{ForwardBatch}, the store backend parses the input token IDs for each prefill and decode batch. The prefetch mechanism is launched asynchronously, allowing the retrieved embeddings to be transferred directly from CXL pool to the GPU memory.

\textbf{Computation:} Each \texttt{ModelRunner} processes its assigned requests or model shards according to the configured parallelism strategy (DP, TP or PP). Since Engram retrieval places a modest load on system bandwidth, each rank retrieves its required embeddings directly from the pool. These embeddings are then computed with each worker's hidden states at the Engram layer.

  
  

\section{Experiments}
\subsection{Experimental Setup}
\label{sec:exp_set}
The experiment is conducted on a testbed consisting of two server nodes, each equipped with dual Intel Xeon 6766E processors, 288 total cores at 1.9 GHz, 216 MiB of L3 cache, and 2 NVIDIA L20 GPUs. Each server features 1 TiB of local memory and runs Ubuntu 22.04.5 LTS. To facilitate memory pooling, the nodes are connected via a XConn XC50256 CXL switch to a 256 GB Sumsung memory card, which utilizes the Montage M88MX5851 CXL memory expansion controller.

\subsection{CXL Engram Read Latency}
\begin{figure}
    \centering
    \includegraphics[width=1.0\linewidth]{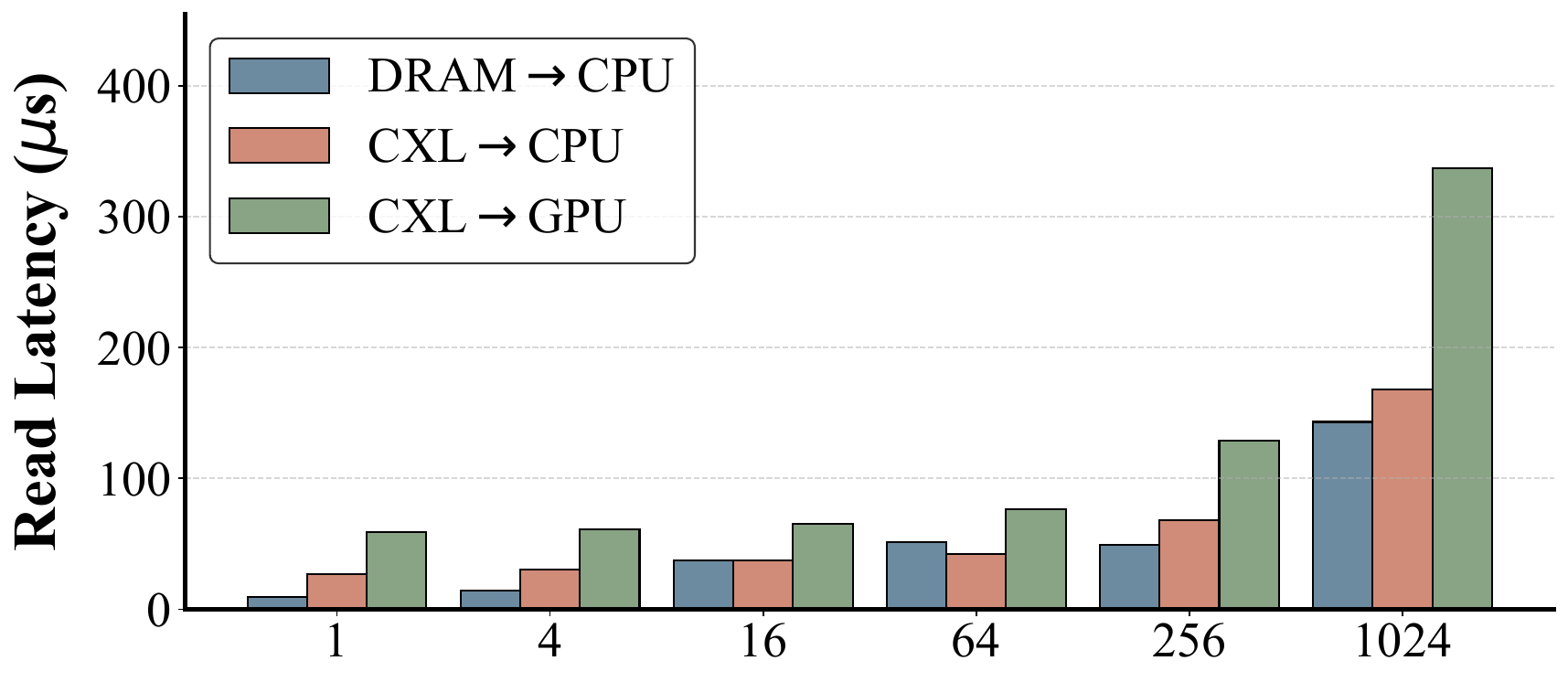}
    \vspace{-20pt}
    \caption{Latency for Engram-27B across varying batch size.}
    \vspace{-10pt}
    \label{fig:27}
\end{figure}

\begin{figure}
    \centering
    \includegraphics[width=1.0\linewidth]{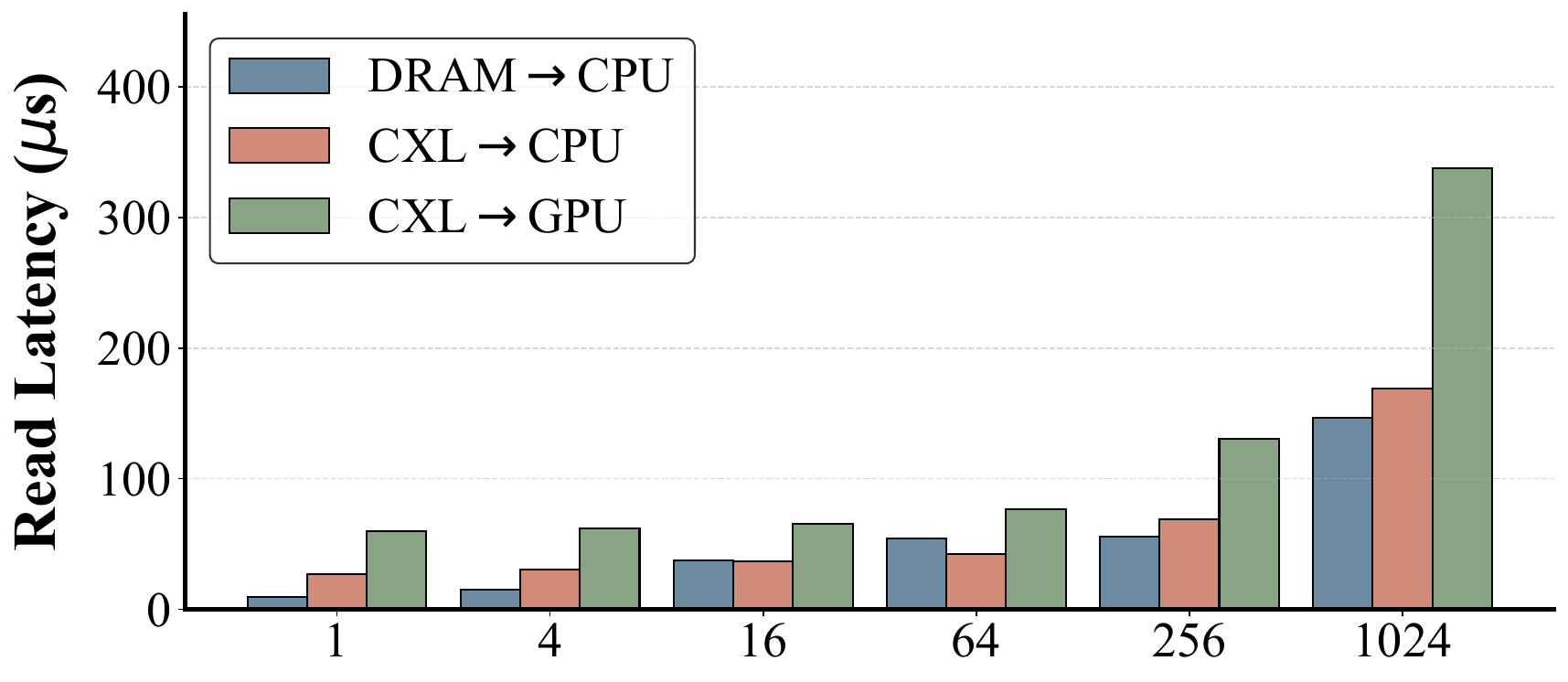}
    \vspace{-20pt}
    \caption{Latency for Engram-40B across varying batch size.}
    \vspace{-10pt}
    \label{fig:40}
\end{figure}

We evaluate the CXL read latency for Engram embeddings under
Engram-27B ($\text{vocab\_size}=2,262,400$; $\text{emb\_dim}=1,280$) and Engram-40B ($\text{vocab\_size}=7,239,680$; $\text{emb\_dim}=1,280$) configurations across varying retrieval batch sizes, with results detailed in Figure~\ref{fig:27} and Figure~\ref{fig:40}.



Experimental data indicates that CXL-to-CPU latency is comparable to local DRAM. While direct CXL-to-GPU reads using custom kernels involve higher latency, the overhead remains within an acceptable range. Furthermore, CXL read efficiency does not diminish as the Engram parameters scale. This stability underscores the feasibility of CXL memory pooling for future high-capacity Engram modules.

\subsection{End-to-end Inference Throughput}
We integrate Engram modules into SGLang~\cite{sglang}. Due to the lack of open-source Engram weights and models, we utilizes Qwen3-4B and Qwen3-8B~\cite{qwen3technicalreport} to emulate the performance. To reflect the overhead, Engram computations are executed during the forward pass and the retrieved embeddings are not added to the hidden states to preserve model stability.

The end-to-end throughput (tokens/s) is summarized in Table~\ref{tab:engram_sgl}. Measurements were conducted on a single L20 GPU (DP=1) with a batch size of 256 and a 512-token input/output sequence length. While the addition of Engram modules introduces a marginal reduction in throughput compared to the baseline, the results demonstrate that CXL-based Engram pools achieve near-DRAM performance. 

\begin{table}[h]
\centering
\caption{Model throughput analysis with Engram.}
\vspace{-10pt}
\label{tab:engram_sgl}
\renewcommand{\arraystretch}{0.9}
\begin{tabular}{ll|c}
\toprule
\textbf{Model} & \textbf{Configuration (DP=1)}  & \textbf{Throughput} \\
\midrule
\multirow{3}{*}{Qwen3-4B} & Baseline  & 6183.9 \\
                          & +Engram (DRAM) & 5683.7 \\
                          & +Engram (CXL) & 5614.4 \\
\midrule
\multirow{3}{*}{Qwen3-8B} & Baseline & 4185.6 \\
                          & +Engram (DRAM) & 3909.7 \\
                          & +Engram (CXL) & 3895.0 \\
\bottomrule
\end{tabular}
\end{table}

Furthermore, we investigate the scalability of system by increasing the Data Parallelism (DP) and the number of nodes (nnode), thereby increasing the stress of CXL pool access. As shown in Table~\ref{tab:exp_scale}, scaling to DP=2 and nnode=2 yields only a negligible performance drop compared to the single-node setup. This indicates that even within our prototype testbed, the CXL-based Engram pool exhibits robust scalability. We anticipate that by using additional CXL controllers and memory interleaving, the architecture can seamlessly scale to larger shared-memory clusters.

\begin{table}[h]
\centering
\caption{Scalability analysis of CXL-based Engram pool.}
\vspace{-10pt}
\label{tab:exp_scale}
\renewcommand{\arraystretch}{0.9}
\begin{tabular}{ll|c}
\toprule
\textbf{Model} & \textbf{Configuration (CXL)}  & \textbf{Throughput} \\
\midrule
\multirow{4}{*}{Qwen3-4B} & +Engram (DP=1)  & 5614.4 \\
                          & +Engram (DP=1, nnode=2) & 5528.3  \\
                          & +Engram (DP=2) & 8180.6 \\
                          & +Engram (DP=2, nnode=2) & 8106.2 \\
\midrule
\multirow{4}{*}{Qwen3-8B} & +Engram (DP=1)  & 3895.0 \\
                          & +Engram (DP=1, nnode=2) & 3883.2  \\
                          & +Engram (DP=2) & 5661.5 \\
                          & +Engram (DP=2, nnode=2) & 5629.7 \\
\bottomrule
\end{tabular}
\end{table}

\subsection{Cost Analysis}

We perform a comparative cost analysis between local DRAM configurations and CXL-based memory pooling. 
In the local DRAM model, each cluster node must be provisioned with the full memory capacity required to host the Engram module (e.g., 200GB for a 100B Engram table). Conversely, the CXL-based architecture utilizes a shared memory pool, allowing multiple nodes to access a single, centralized memory resource. 
Table~\ref{tab:component_costs} details the estimated costs for essential hardware components~\cite{yang2025beluga}. Given the persistent rise in memory pricing~\cite{Counterpoint2026Memory}, the economic utility of CXL-based memory pooling has become increasingly pronounced.

\begin{table}[h]
\centering
\begin{threeparttable}
\captionsetup{skip=5pt}
\caption{Hardware cost of DRAM and CXL pool.}
\label{tab:component_costs}
\begin{tabular}{l|l|r}
\hline
\textbf{Component} & \textbf{Specification} & \textbf{Cost} \\ \hline
DDR5 RDIMM & DRAM (per GB) & \$15.00 \\ \hline
CXL Switch & XConn, 32$\times$PCIe 5.0 x16 & \$5,800.00 \\ \hline
CXL Adapter & PCIe/CXL x16 Card & \$210.00 \\ \hline
CXL Controller & Memory Expansion ASIC & \$300.00 \\ \hline
\end{tabular}
\begin{tablenotes}
    \scriptsize
    \item * B1 sample price, and final price subject to change.
\end{tablenotes}
\end{threeparttable}
\end{table}

Based on the estimated unit cost, Table \ref{tab:cost_comparison} presents a comparative analysis of the total capital expenditure. We assume each host node is equipped with a CXL host adapter, pairing with a dedicated CXL controller within the memory pool. Our projections indicate that while CXL entails substantial fixed infrastructure costs, resulting in higher costs for small-scale configurations (e.g., 100B Engram across two nodes), it exhibits superior cost-efficiency as the system scales in both node count and model complexity. 


\begin{table}[htbp]
\centering
\caption{Engram storage cost comparison.}
\vspace{-10pt}
\label{tab:cost_comparison}
\renewcommand{\arraystretch}{0.8} 
\begin{tabular}{@{}c|c|rrr@{}}
\toprule
\textbf{Engram} & \textbf{Nodes} & \textbf{Local} & \textbf{CXL Pool} & \textbf{Savings} \\ \midrule
\multirow{4}{*}{100B} & 2  & \$6,000  & \$9,820  & -\$3,820\\ 
                      & 4  & \$12,000 & \$10,840 & \$1,160 \\  
                      & 8  & \$24,000 & \$12,880 & \$11,120 \\ 
                      & 16 & \$48,000 & \$16,960 & \$31,040 \\ \midrule
\multirow{4}{*}{400B} & 2  & \$24,000 & \$18,820 & \$5,180 \\ 
                      & 4  & \$48,000 & \$19,840 & \$28,160 \\ 
                      & 8  & \$96,000 & \$21,880 & \$74,120 \\ 
                      & 16 & \$192,000 & \$25,960 & \$166,040 \\ \bottomrule
\end{tabular}
\end{table}
\section{Discussion}
Extending beyond the CXL-based implementation, we discuss two critical aspects for further enhancing the practicality of Engram pooling in this section.

\textbf{RDMA Potential in Engram. } In our experiment in Section~\ref{sec:rdma_cxl}, we employ Mooncake Store~\cite{mooncake} as a direct RDMA backend for Engram storage. However, we observe performance bottlenecks due to centralized management and the inefficiency of \texttt{get\_batch} when handling discrete, small packets. To address this, it is necessary to design a direct P2P transfer interface in Mooncake Store and aggregate small data payloads prior to RDMA transmission. Additionally, caching "hot" Engram embeddings in DRAM can further enhance performance. As RDMA is a more ubiquitous hardware standard, we believe that it can also be effective for Engram if supported by a custom-tailored implementation.

\textbf{Coexistence with KV Cache. } Traditionally, the memory pool was dedicated solely to the KV Cache in LLM. Engram introduces a novel storage and access pattern within this shared pool. Effectively managing memory allocation and mitigating potential access conflicts between these distinct workloads remains an open research challenge.

\section{Conclusion}
This paper introduces the first CXL-based system to offload Engram parameters to a shared memory pool. We showcase that the memory access pattern of Engram is highly compatible with CXL, enabling near-DRAM end-to-end performance on inference framework. 
With the CXL-based pool, our approach significantly reduces the deployment costs of Engram without sacrificing inference performance. This paves the way for more scalable and cost-efficient memory expansion in next-generation LLM infrastructures.


\newpage

\bibliographystyle{ACM-Reference-Format}
\bibliography{references}

\end{document}